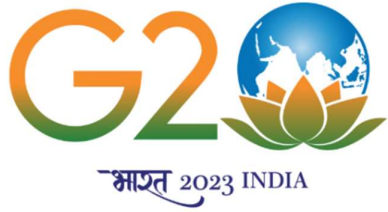 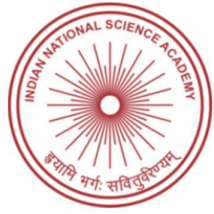 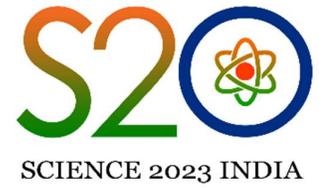

# Policy Brief

## REGIONAL AND GLOBAL COLLABORATIONS IN ASTRONOMY

Discussed and drafted during S20 Policy Webinar on Astroinformatics for Sustainable Development held on 6-7 July 2023

Contributors: Kazuhiro Sekiguchi, Jess McIver, Annapurni Subramaniam, Eswar Reddy, R Srianand, Reinaldo Rosa, Stefano Andreon, Tarun Souradeep, Bernard Fanaroff, Rafael Santos, Paula Coelho, Pranav Sharma, Ashish Mahabal

**Introduction**
Astronomy brings together advanced scientific research, state-of-the-art technology, and educational initiatives, all while captivating and stimulating people of all ages. By doing so, it possesses the potential to serve as a powerful catalyst for sustainable global development and the resolution of global societal issues. It attracts a diverse range of scientists and experts from various fields, fostering collaboration and innovation.

By leveraging their resources, influence, and diplomatic initiatives, S20 academies can foster an enabling environment for international collaborations in astronomy, facilitate knowledge exchange, and drive scientific advancements that benefit humanity. This policy brief explores the opportunities and challenges presented by regional and global collaborations in astronomy.

**Opportunities**
Regional and global collaborations in astronomy offer a wide range of opportunities to advance scientific knowledge, foster international cooperation, and contribute to the betterment of society through scientific discoveries, technological advancements, and public engagement. Some of these opportunities include:

1. **Sharing knowledge and expertise**

    **Data Sharing and Knowledge Exchange:** Collaboration in astronomy fosters knowledge sharing and education, particularly in STEM fields. It enables the exchange of expertise, best practices, and research findings, allowing scientists and educators to learn from one another. This knowledge dissemination and transfer contribute to developing skilled scientists and researchers worldwide. Additionally, astronomy's inherent interdisciplinary nature encourages cross-disciplinary collaboration, which can lead to innovative solutions for sustainable development challenges.

    **Access to Diverse Expertise:** Collaborations bring together scientists from different regions, institutions, and backgrounds, creating opportunities for interdisciplinary research and access to diverse expertise. This enhances the quality of research and fosters innovative approaches to solving complex astronomical problems.

2. **Resources and Infrastructure Sharing**

    **Shared Access to Facilities:** Collaborations facilitate shared access to astronomical facilities and resources. Researchers from participating regions can access telescopes, observatories, and data archives they might not have in their home countries, broadening their observational capabilities and enabling new scientific investigations.

    **Large-scale projects and cost-sharing:** Regional and global collaborations enable large-scale projects that require significant resources and infrastructure. Examples include building and operating large telescopes, observatories, and space missions. These projects can lead to groundbreaking discoveries and advancements in our understanding of the universe.

3. **Training and Capacity Building**
    Collaborations provide training and capacity-building opportunities, especially for scientists and students from regions with limited resources. Participating researchers can gain knowledge, skills, and exposure to advanced research techniques through

collaborations with more established astronomical institutions, fostering the development of scientific expertise globally.

4. **Global Networking**
   Regional and global collaborations create extensive networks of scientists, fostering long-term relationships and partnerships. These networks enable the exchange of ideas, sharing of best practices, and joint research efforts that lead to accelerated scientific progress and discoveries.

5. **Technological Advancements and Innovation**
   Collaborations encourage technological advancements and innovation in astronomy. Researchers from different regions can share their expertise in instrumentation, data analysis techniques, and computational methods, leading to the development of new technologies and tools that benefit the entire astronomical community.

6. **Public Engagement and Outreach**
   Collaborations provide opportunities for joint public engagement and outreach activities. By working together, researchers can enhance public awareness of astronomy, inspire young minds, and foster a sense of global scientific community. Collaborations can also contribute to cultural exchange and promote international goodwill.

7. **Addressing Global Challenges**
   Regional and global collaborations in astronomy can contribute to addressing global challenges such as climate change, space debris monitoring, and planetary defence. Collaborations can contribute to scientific understanding and inform decision-making by sharing data and expertise.

**Challenges**

Despite the numerous opportunities offered by regional and global cooperation, some challenges must be addressed. These challenges include issues of communication and language, cultural differences, time zone differences, as well as:

- **Funding and Resource Allocation:** Collaborations often require financial resources for infrastructure, equipment, and research activities. Securing funding from multiple sources and ensuring equitable resource allocation can be complex. Different countries or regions may have varying levels of financial support and priorities, posing challenges to sustaining collaborations.

- **Intellectual Property and Data Sharing:** Collaborative research in astronomy involves sharing data, software, and intellectual property. Establishing agreements on data sharing, access, and intellectual property rights can be challenging, as countries may have different policy frameworks and regulations.

- **Data Access and Security:** Collaborations rely on the sharing and analysing astronomical data, which may be subject to data access restrictions or security concerns. Ensuring data privacy, cybersecurity and addressing legal and ethical considerations can pose challenges in collaborative projects.

- **Technology and Infrastructure:** Collaborations in astronomy often rely on advanced technology and infrastructure, such as telescopes, observatories, and data centres. Ensuring compatibility and interoperability of different systems can be challenging, especially when collaborators use other technologies or have varying levels of infrastructure development.

- **Governance and Decision-making:** Collaborations involving multiple regions or countries need effective governance structures and decision-making processes. Balancing the interests and priorities of different stakeholders and ensuring inclusive participation can be challenging, particularly in large-scale global collaborations.

- **Unequal Capacities and Expertise:** Collaborations may involve partners with varying scientific expertise, infrastructure, and research capacities. Bridging the gaps between different regions or countries regarding skills, resources, and knowledge can be challenging, requiring capacity-building initiatives and technology transfer programs.

Addressing these challenges requires proactive efforts to establish effective communication channels, foster trust, develop shared protocols and policies, and maintain flexibility in accommodating diverse needs and perspectives.

**Opportunities for Collaboration**
The S20 science academies can play a significant role in overcoming the challenges in regional and global collaborations in astronomy. Here are some ways they can contribute:

- **Data Sharing and Open Science:** S20 science academies can promote open science and data-sharing practices in astronomy by encouraging the development of data repositories, open-access databases, and platforms that facilitate the sharing and dissemination of research findings and datasets. In addition, they can support initiatives that ensure the ethical and responsible use of shared data. To achieve this, they can help their governments establish data-sharing policies and protocols that address privacy and security concerns, implement secure data management systems following cybersecurity best practices, and develop ethical frameworks and guidelines for the responsible use of shared data. By taking these steps, S20 academies can foster a culture of open collaboration and maximise the scientific potential of astronomical research.

- **Funding and Support:** S20 academies can provide counsel to their governments about allocation of financial resources to support international collaborations in astronomy by establishing dedicated funds or grant programs for global astronomical research initiatives. By providing substantial funding, governments can effectively address resource gaps and ensure the long-term sustainability of collaborative projects. In addition to this, transparent and equitable mechanisms for resource allocation should be established, and collaboration among funding agencies should be encouraged to develop joint funding opportunities that specifically support international projects in the field of astronomy.

- **Policy and Regulatory Frameworks:** G20 governments should foster international cooperation in astronomy by implementing favourable policies and regulatory frameworks. To promote smooth collaboration, governments must develop explicit agreements on key aspects like data sharing, intellectual property rights, technology transfer, and people-to-people exchanges. By doing so, they can ensure that all parties involved know their rights and responsibilities from the outset.

- **Science Diplomacy:** G20 governments can leverage their diplomatic channels and engage in science diplomacy efforts to strengthen international collaborations in astronomy. They can promote collaboration, exchange programs, and joint research initiatives by fostering diplomatic relations and scientific cooperation agreements with other nations. To encourage exchanges of researchers among the G20 countries, requirements such as visas should be relaxed.

- **Capacity Building and Training:** Through capacity-building programs, G20 governments could bolster scientific expertise and infrastructure in countries with limited resources. To achieve this, scholarships, grants, and training opportunities for scientists from developing regions can be offered, enabling them to collaborate with leading astronomical institutions within G20 countries and acquire valuable skills and knowledge. By fostering collaboration in training and capacity-building initiatives, facilitating the exchange of knowledge and expertise, developing mentorship programs, and encouraging scientific exchanges, governments can promote skill development and facilitate technology transfer among partners.

- **Technology and Infrastructure Sharing:** S20 academies can support and provide expertise to their governments to inculcate opportunities that facilitate collaboration and technology sharing among participating countries, leading to the development and dissemination of advanced technologies. S20 academies can support their governments to encourage collaborative efforts to establish common standards and protocols, ensuring regional compatibility and interoperability. Additionally, S20 academies can provide expertise to their governments to support technology transfer initiatives, bridging technological gaps and enhancing the capabilities of less-developed regions. By fostering collaboration, establishing standards, and promoting infrastructure and expertise sharing, S20 academies can help their governments to drive technological advancements and resource availability for all benefits.

- **Education and Public Outreach:** S20 academies can encourage dialogue with their governments for investment in public outreach programs that promote science education, astronomy awareness, and engagement with the public. By fostering an interest in astronomy, they can create a supportive environment for collaboration and encourage young talents to pursue careers in science and technology.

- **International Cooperation Frameworks:** S20 academies and the International Astronomical Union (IAU) should collaborate to establish international frameworks for cooperation in astronomy[1]. They can work together to develop guidelines, standards, and protocols that promote collaboration, address common challenges, and facilitate knowledge exchange among participating nations.

- **Policy Coordination:** S20 academies can support should engage in policy coordination efforts to align their strategies and approaches to support their governments establish regional and global collaborations in astronomy. By sharing best practices and experiences, they can learn from one another and collectively address challenges that transcend national boundaries.

---

[1] IAU Strategic Plan 2020-2030. (2018). International Astronomical Union | IAU. https://www.iau.org/administration/about/strategic_plan/

By leveraging their resources, influence, and diplomatic channels, S20 academies can support their governments to foster an enabling environment for international collaborations in astronomy, facilitate knowledge exchange, and drive scientific advancements that benefit humanity.

**Conclusion**
The large amount of data obtained by modern astronomical observational research not only solves the mysteries of the universe but also serves as an intellectual property shared by humanity. Regional and global collaborations in astronomy present significant opportunities for advancing scientific understanding and promoting global development. While several challenges must be addressed, we believe the recommendations outlined in this white paper can help promote more effective and inclusive collaborations in astronomy research. We urge policymakers, funding agencies, and other stakeholders to support these recommendations and promote regional and global partnerships in astronomy.

**S20 Co-Chair**: Ashutosh Sharma, Indian National Science Academy
**INSA S20 Coordination Chair:** Narinder Mehra, Indian National Science Academy

**Contributors**
Kazuhiro Sekiguchi, National Astronomical Observatory of Japan, Japan
Jess McIver, University of British Columbia, Canada
Annapurni Subramaniam, Indian Institute of Astrophysics, India
Eswar Reddy, Indian Institute of Astrophysics, India
R Srianand, Inter-University Centre for Astronomy and Astrophysics, India
Reinaldo Rosa, Instituto Nacional de Pesquisas Espaciais-INPE, Brazil
Stefano Andreon, INAF-OA Brera, Italy
Tarun Souradeep, Raman Research Institute, India
Bernard Fanaroff, South African Radio Astronomy Observatory, South Africa
Rafael Santos, Instituto Nacional de Pesquisas Espaciais, Brazil
Paula Coelho, University of São Paulo, Brazil
Pranav Sharma, Indian National Science Academy, India
Ashish Mahabal, California Institute of Technology, USA

5